\documentclass[final,5p,times,twocolumn,num]{elsarticle}

\usepackage{amssymb}
\usepackage{csquotes}
\usepackage{pgf}
\usepackage[separate-uncertainty]{siunitx}
\usepackage{physics}
\usepackage{cleveref}
\usepackage{mathtools}

\journal{Physics Letters B}

\begin{document}

\begin{frontmatter}


\title{Elastic deuteron-deuteron scattering in the ${}^5S_2$ channel within Nuclear Lattice Effective Field Theory}

\author[first]{Helen Meyer}
\affiliation[first]{organization={Helmholtz Institut für Strahlen- und Kernphysik and
            Bethe Center for Theoretical Physics,
            Universität Bonn},
            city={Bonn},
            postcode={D-53115},
            country={Germany}}

\author[second]{Serdar Elhatisari}
\affiliation[second]{organization={Faculty of Engineering and Natural Sciences, Gaziantep Islam Science and Technology University},
            city={Gaziantep},
            postcode={27010}, 
            country={Türkiye}}

\author[third]{Fabian Hildenbrand}            
\affiliation[third]{organization={Institute for Advanced Simulation (IAS-4), Forschungszentrum Jülich},
            city={Jülich},
            postcode={D-52425}, 
            country={Germany}}

\author[first,third,fourth]{Ulf-G.~Meißner}
\affiliation[fourth]{organization={Peng Huanwu Collaborative Center for Research and Education,
International Institute for Interdisciplinary and Frontiers, Beihang University},
            city={Beijing},
            postcode={100191}, 
            country={China}}


\begin{abstract}
We calculate low-energy deuteron-deuteron scattering in the spin-quintet $^{5}S_2$ channel using nuclear lattice effective field theory. The calculation combines chiral interactions at next-to-next-to-next-to-leading order, implemented through wavefunction matching, with the adiabatic projection method. Because the radial cluster basis develops small norm-matrix eigenvalues at large Euclidean projection time, we investigate two stabilization procedures: Tikhonov regularization and projection onto well-resolved norm eigenmodes. The two procedures yield consistent Coulomb-subtracted phase shifts within their statistical and numerical uncertainties. A Coulomb-modified effective-range analysis gives ${}^5a_{dd} = \SI{11.99 \pm 0.22}{\femto\m}$ and ${}^5r_{dd} = \SI{3.44 \pm 0.79}{\femto\m}$. The phase shifts are more negative, and the scattering length is substantially larger than in previous calculations, corresponding to a stronger effective repulsion in the $^{5}S_2$ channel. These results provide a first nuclear-lattice benchmark for deuteron-deuteron scattering and establish a starting point for future coupled-channel calculations of the deuteron-induced reactions relevant to big-bang nucleosynthesis.
\end{abstract}


\end{frontmatter}



\section{Introduction}
\label{introduction}

Big bang nucleosynthesis (BBN) (for reviews see e.g.~\cite{Wagoner:1966pv, Cyburt:2015mya}) works as an excellent probe of physics beyond the standard model as it is the first process in the history of the early universe that is well understood. Measurements for primordial element abundances are getting more and more precise. In order to have reliable theoretical predictions for BBN that can be compared to observations it is crucial to know the dominant reaction rates precisely. In \cite{Pitrou:2021vqr}, the significance of the deuteron-deuteron reactions $d(d,n){}^3\mathrm{He}$ and $d(d,p){}^3\mathrm{H}$ for the accuracy of theoretical BBN predictions was laid out. Similarly, Fig.~3 in \cite{Burns:2023sgx} illustrates the so-called \enquote{deuteron anomaly}: depending on the source of the twelve most important reaction rates of BBN (NACRE II \cite{Xu:2012uw} or the rates used in the \texttt{PRIMAT} code \cite{Pitrou:2018cgg}), the deuterium primordial abundance either lies within the error bands of the measurement or not. Finding an accurate theoretical description of the key reaction rates in BBN would therefore be the reasonable next step in further understanding BBN and improving the accuracy of theoretical predictions for primordial abundances.

Within the framework of nuclear lattice effective field theory (NLEFT, for a detailed introduction see \cite{Lahde:2019npb}), one can calculate the reaction rates for $d(d,n){}^3\mathrm{He}$ and $d(d,p){}^3\mathrm{H}$ ab initio. As a first step towards calculating these rates, we consider the single-channel $d+d$, ${}^{5}S_2$ scattering problem in NLEFT. Because the deuteron is weakly bound and its wave function thus broad, treating two deuterons on a confined lattice is demanding and requires a careful analysis. Elastic deuteron-deuteron phase shifts have first been calculated in \cite{Meier:1975ydi} using the resonating group method (RGM) \cite{Wildermuth:RGM1,Wildermuth:RGM2} with a Malfliet-Tjon two-nucleon potential \cite{Malfliet:1968tj} (MT I and MT III).  The same potential was used in \cite{Filikhin:2000}, where the four-nucleon Yakubovskii equations were instead solved with the cluster-reduction method \cite{Yakovlev:1997vi}. In \cite{Hofmann:1996jv}, results from calculating $d+d$ scattering phase shifts with the RGM using a Bonn-based Gaussian potential \cite{Kellermann:1989ay} were compared to an $R$-matrix analysis of experimental data of the $^{4}\mathrm{He}$ system. For the $S=0$ and $S=2$ $S$-wave phase shifts both approaches gave very similar results. This analysis was further refined in \cite{Hofmann:2005iy}, where the Argonne v18 two-nucleon potential \cite{Wiringa:1994wb} and the Urbana IX three-nucleon potential \cite{Pudliner:1997ck} were used with the RGM. The most recent calculation of elastic $d+d$ phase shifts was performed in \cite{Carew:2021jai}: a variational-bound formulation based on the Faddeev–Yakubovskii1 chain-of-partition many-particle equations \cite{Carew:2014} was used together with a nucleon-nucleon Gaussian potential. There are significant differences among the results obtained in these works, which provides additional motivation for an ab initio lattice calculation of the $d+d$ system. In addition, the $S=2$ scattering length and effective range for elastic $dd$ scattering were calculated in Ref.~\cite{Rojik:2025jcv} in pionless EFT up to NLO.
The physical four-nucleon problem in the $J^\pi=2^+$ sector is, however, a multichannel problem. In addition to the $d+d$, ${}^{5}S_2$ channel, it contains $d+d$ $D$-wave and $G$-wave channels
and the open $p+{}^3\mathrm{H}$ and $n+{}^3\mathrm{He}$ rearrangement channels (${}^{1}D_2$, ${}^{3}D_2$). The treatment of this channel structure differs among the previous calculations. Refs.~\cite{Hofmann:1996jv,Hofmann:2005iy} performed explicit coupled-channel analyses, whereas the other low-energy calculations quoted above primarily report an $S$-wave-projected $d+d$ phase shift. Ref.~\cite{Carew:2021jai} also considered three- and four-body breakup scattering at higher energies. Full coupled-channel calculations for the four-nucleon system for many partial waves have also been done in Refs.~\cite{Deltuva:2007xw,Deltuva:2010yp,Arai:2011zz,Deltuva:2015fta,Deltuva:2015mfa,Viviani:2022qfy,Solovyev:2024fmp}. However, the main focus of these works was the calculation of observables for the $d(d,n){}^3\mathrm{He}$ and $d(d,p){}^3\mathrm{H}$ reactions. Only Refs.~\cite{Deltuva:2007xw,Deltuva:2015fta,Deltuva:2015mfa} give explicit results for elastic deuteron-deuteron scattering. In these papers, the differential cross section as a function of the scattering angle for some selected deuteron energies  was calculated including partial waves up to $J=3$. A direct comparison between our single-channel calculation and a differential cross section incorporating contributions from multiple partial waves is not particularly meaningful.

In the present first study, we restrict the asymptotic adiabatic basis to the $d+d$ ${}^{5}S_2$ channel. The reported phase shift should therefore be understood as the phase shift of the projected single-asymptotic-channel Hamiltonian. It is not a determination of the complete coupled-channel $J^\pi=2^+$ scattering matrix, its inelasticities, or its coupled-channel eigenphases, as the computational costs required for a full analysis including all eight coupled channels are beyond the available resources. Being the channel with the lowest orbital angular momentum, we expect the $dd$ ${}^{5}S_2$ channel to be the dominant contribution at low energies.

NLEFT is a lattice formulation of Chiral Effective Field Theory (ChEFT, for a detailed review see \cite{Epelbaum:2008ga}) where nucleons (neutrons and protons) are the relevant particle degrees of freedom. These are put on a cubic lattice with periodic boundary conditions and subjected to the ChEFT potential at a given order. Observables are then computed using advanced Monte Carlo methods (like auxiliary-field quantum Monte Carlo). With the so-called \enquote{wavefunction matching} method \cite{Elhatisari:2022zrb} it has become possible to calculate observables accurately at N$^3$LO (next-to-next-to-next-to leading order of the chiral expansion). Its recent successes include the precise ab initio calculation of the properties of heavy nuclei \cite{Hildenbrand:2025voq}, precise nuclear radii \cite{Ren:2025vpe}, scattering phase shifts \cite{Elhatisari:2015iga,  Li:2018ymw,Elhatisari:2021eyg,Elhatisari:2025fyu} and even properties of dense neutron stars \cite{Tong:2024jvs} (using a simplified action). For a recent review, see~\cite{Lee:2025req}, and $\alpha$--$\alpha$ scattering at N$^3$LO is discussed in Ref.~\cite{Sarkar:2026wev}.

Our work is structured as follows: In \cref{s:NLEFT} we briefly lay the
methods underlying our calculations. The results are presented and discussed in
\cref{s:results}. The Appendix contains some technicalities related
to the extraction of the deuteron-deuteron phase shifts.


\section{NLEFT and the Adiabatic Projection Method}\label{s:NLEFT}

In this work, we employ the wavefunction matching method introduced in Ref.~\cite{Elhatisari:2022zrb} to calculate the elastic deuteron-deuteron phase shift in the spin-quintet ${}^{5}S_{2}$ channel using chiral interactions at N$^3$LO. The theoretical framework follows that described in detail in Ref.~\cite{Elhatisari:2025fyu}. We denote the original high-fidelity chiral Hamiltonian by
\begin{equation}
H^{\mathrm{N3LO}}
=
K
+V_{\mathrm{OPE}}
+V_{\mathrm{C}}
+V_{3N}^{Q^3}
+V_{2N}^{Q^4}
+W_{2N}^{Q^4},
\label{eq:HN3LO}
\end{equation}
where $K$ is the kinetic-energy operator, $V_{\mathrm{OPE}}$ is the one-pion-exchange potential including its short-range counterterm, and $V_{\mathrm{C}}$ is the Coulomb interaction. The term $V_{3N}^{Q^3}$ contains the three-nucleon interactions entering at N$^2$LO in the chiral expansion, while $V_{2N}^{Q^4}$ denotes the short-range two-nucleon interactions at N$^3$LO. The term $W_{2N}^{Q^4}$ restores Galilean invariance for the two-nucleon interactions on the lattice. Details of the Coulomb interaction and the two-nucleon forces are given in Refs.~\cite{Li:2018ymw,Elhatisari:2022zrb}.

Wavefunction matching applies a finite-range unitary transformation,
\begin{equation}
H'^{\mathrm{N3LO}}
=
U^\dagger H^{\mathrm{N3LO}}U,
\label{eq:Hmatched}
\end{equation}
where $U$ acts nontrivially only when the separation of two nucleons is smaller than a matching radius $R$. The transformation is constructed such that, in each relevant two-nucleon channel, the low-energy wave functions of $H'^{\mathrm{N3LO}}$ are proportional to those of a simple Hamiltonian $H_S$ for separations $r<R$, while they coincide with the wave functions of the original high-fidelity Hamiltonian for $r>R$. The simple Hamiltonian $H_S$ consists of the kinetic-energy operator, a short-range smeared SU(4)-symmetric two-nucleon contact interaction, and the one-pion-exchange potential.

We write the transformed Hamiltonian as
\begin{equation}
H'^{\mathrm{N3LO}}
=
H_S+\Delta H,
\qquad
\Delta H
=
H'^{\mathrm{N3LO}}-H_S.
\label{eq:Hpartition}
\end{equation}
The simple Hamiltonian $H_S$ is treated nonperturbatively in the Euclidean-time projection, whereas the correction $\Delta H$ is included through first order in perturbation theory. Since the unitary transformation makes $H'^{\mathrm{N3LO}}$ considerably closer to $H_S$ than the original high-fidelity Hamiltonian, the perturbative expansion converges rapidly, while the use of $H_S$ for the nonperturbative evolution strongly suppresses Monte Carlo sign oscillations.

\begin{figure}
    \centering
    \includegraphics{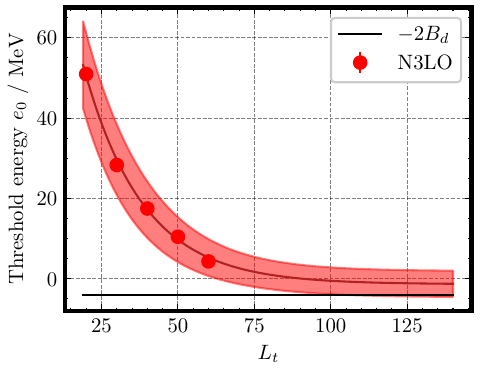}
    \caption{Threshold energy of two non-interacting deuterons on the lattice as a function of Euclidean time $L_t$. The red band shows the result from an exponential fit as in \cref{eq:e0Ltinf}, the black line is the expected result of twice the deuteron binding energy at N$^3$LO.}
    \label{fig:e0_Ltinf}
\end{figure}

The expectation value of an operator is
\begin{equation}
    \expval{O} = \frac{\mel{\Psi_I}{M^{L_t/2}OM^{L_t/2}}{\Psi_I}}{\mel{\Psi_I}{M^{L_t}}{\Psi_I}},
\end{equation}
where $L_t$ denotes the number of temporal lattice-steps and $M$ is the transfer matrix $:\exp(-H_S \alpha_t):$, with $\alpha_t$ the ratio of temporal to spatial lattice spacing. The initial many-body states $\Psi_I$ are Slater determinants constructed from single-particle orbitals localized around the two cluster centers. For each proton-neutron cluster, we use a compact radial trial profile inspired by the asymptotic $S$-wave deuteron wave function,
\begin{equation}\label{eq:init_wf}
    \Psi(r) = \frac{1}{r}e^{-12.5 \sqrt{m_N \, B_d}r}.
\end{equation}
Here, $m_N=\SI{938.92}{\MeV}$ is the average nucleon mass and $B_d=\SI{2.224}{\MeV}$ is the deuteron binding energy. In the numerical implementation, all quantities in the exponent are expressed in consistent lattice units. The value at zero relative separation is defined to be $1/(0.5\,\text{l.u.})$. The factor of $12.5$ is tuned to ensure that the initial wave function is not too broad and there are no instabilities coming from box size effects when putting two deuterons on the lattice. 
These expectation values are calculated for a number of different $L_t$ (we calculate $L_t = 20, 30, 40, 50$ and $60$ here) and then extrapolated to $L_t\to\infty$. This evolution in Euclidean time filters out the lowest energy states of the system.
We determine that the threshold energy obtained by simulating two non-interacting deuterons on the lattice is consistent within uncertainties with twice the true deuteron ground state energy when extrapolating to infinite Euclidean time. This is illustrated in \cref{fig:e0_Ltinf}: we fit an exponential
\begin{equation}\label{eq:e0Ltinf}
    e_0(L_t) = E_\infty + ce^{-\Delta E L_t \alpha_t}
\end{equation}
to the threshold energy as a function of $L_t$ and obtained $E_\infty = \SI{-1.331\pm 3.263}{\MeV}$. The rather large uncertainty reflects the $\chi^2_\mathrm{red}$ value of the fit and the fact that the data has not converged to a constant yet for $L_t = 60$ and so $E_\infty$ cannot be so precisely determined. The threshold energy $e_0(L_t)$ shown in \cref{fig:e0_Ltinf} is obtained from the zero relative-momentum component of the Hamiltonian for two non-interacting, Euclidean-time-dressed deuterons. It is a finite $L_t$ energy estimator, rather than a converged determination of twice the deuteron ground-state energy. As described above, the initial cluster states use the compact radial profile in \cref{eq:init_wf}. At the Euclidean times considered, contributions from excited internal states can remain. Therefore, a positive value of $e_0(L_t)$ does not by itself imply that the underlying Hamiltonian produces an unbound or weekly bound deuteron. As will be explained in section~\ref{s:phase_shifts}, the scattering phase shift is extracted from the wave function and not from the energy spectrum and so the convergence behaviour is very different. From the figures in the appendix, it is evident that the phase shift is already converged for $L_t = 60$ and so we constrain the calculation to these Euclidean times.

In this work we calculate on a lattice with spacing $a = \SI{1.316}{\femto\m}$, corresponding to a momentum cutoff of $\Lambda = \pi/a \approx 470\,{\rm MeV}$. We choose a rather large box of $L=\SI{18.4}{\femto\m}$ to minimize finite-volume effects. The temporal lattice spacing is $a_t = \SI{0.20}{\femto\m}$. 

\subsection{Adiabatic Projection Method}

In order to increase the efficiency of the calculation we employ the so-called adiabatic projection method~\cite{Pine:2013zja}, which was used with great success in previous scattering calculations, e.g.~in Refs.~\cite{Elhatisari:2015iga,Elhatisari:2021eyg,Elhatisari:2016hby,Elhatisari:2019fvk}. We construct the initial two-cluster states from two deuteron clusters. Each cluster contains one spin-up proton and one spin-up neutron and therefore has deuteron spin projection $m_d=+1$. For a separation vector, we define
\begin{equation}
\left|\vec{x};M_S=2\right\rangle
= 
\sum_{\vec{n}}
\left|d,m_d=1;\vec{n}+\vec{x}\right\rangle_1
\otimes
\left|d,m_d=1;\vec{n}\right\rangle_2 ,
\label{eq:polarized-cluster-state}
\end{equation}
where the sum is over all lattice sites $\vec{n}$. Because the two spin-one clusters are both maximally polarized,
\begin{equation}
\left|1,1\right\rangle_1
\otimes
\left|1,1\right\rangle_2
= \left|S=2,M_S=2\right\rangle ,
\end{equation}
the initial state belongs uniquely to the spin-quintet channel.

We project the relative motion onto spherical harmonics according to
\begin{equation}
|\Vec{x};S=2,M_S=2\rangle^{\ell m_\ell} = \sum_{\Vec{x}'} Y_{\ell m_\ell}(\hat{x}')\, \delta_{|\Vec{x}|,|\Vec{x}'|}\, |\Vec{x}';M_S=2\rangle.
\label{eq:partial-wave-projection}
\end{equation}
In the present calculation, we take $\ell=0$ and $m_\ell=0$. Coupling the relative orbital angular momentum  $\ell=0$ to $S=2$ therefore gives $J=2$ and $M_J=2$, corresponding to the asymptotic ${}^{5}S_2$ channel. The radial-bin states are then constructed as
\begin{equation}\label{eq:cluster_states}    
\ket{R;M_J=2}^{\ell = 0, m_\ell = 0} = \sum_{|\Vec{x}-\Vec{R}| < a_R/2} \, |\Vec{x};S=2,M_S=2\rangle^{\ell = 0, m_\ell = 0}
\end{equation}
for a given bin size $a_R$. We use $a_R = \SI{0.92}{\femto\m}$ in this work. \cref{eq:partial-wave-projection,eq:cluster_states} specify the initial and asymptotic cluster states in the ${}^{5}S_2$ channel. The Euclidean-time evolution is generated by multiplying with the transfer matrix $M$ defined above. The Euclidean-time-dressed cluster states 
\begin{equation}
    \ket{R;M_J=2}^{\ell = 0, m_\ell = 0}_{n_t} = M^{n_t} \ket{R;M_J=2}^{\ell = 0, m_\ell = 0},
\end{equation}
with $n_t = L_t/2$, are not restricted to a pure $\ell=0$ configuration in the interaction region. In particular, the tensor part of the one-pion-exchange interaction, that is included non-perturbatively in $H_S$, generates the internal $D$-wave component of each deuteron and allows intermediate higher-orbital-angular-momentum components of the four-nucleon state. The remaining spin-dependent interactions contained in $\Delta H$ are included perturbatively.

In the present work, only the asymptotic $\ell=0$, $S=2$ radial channel is retained explicitly in the adiabatic basis. This restriction does not imply that the microscopic Euclidean-time evolution is
confined to two undeformed deuterons in a pure relative $S$ wave. The four nucleons evolve with the microscopic Hamiltonian, so polarization,
deformation, tensor-induced higher-orbital-angular-momentum components, and short-distance configurations with different cluster character can
contribute to the dressed-state matrix elements. However, no independent $p+{}^3\mathrm{H}$, $n+{}^3\mathrm{He}$, or asymptotic $\ell=2,4$
$d+d$ states are included. Consequently, the present basis does not impose scattering boundary conditions in these channels
and cannot determine their outgoing amplitudes, $D$- and $G$-wave phase shifts, or coupled-channel mixing parameters. 

The Hamiltonian $H$ on the lattice is defined in the basis of the dressed cluster states. Because these cluster states are in general non-orthogonal, we define a norm matrix
\begin{equation}
    [N_{L_t}]_{RR'}^{\ell m_\ell} = \prescript{\ell m_\ell}{n_t}{\langle R|R'\rangle}^{\ell m_\ell}_{n_t}.
\end{equation}
The adiabatic Hamiltonian is then (skipping all indices)
\begin{equation}
    H_\mathrm{adiabatic} = N^{-1/2}_{L_t} H N^{-1/2}_{L_t},
\end{equation}
where $H$ is here the sum of all expectation values of operators in the full Hamiltonian, with respect tothe Euclidean-time-dressed cluster states. 

\subsection{Norm matrix regularization}\label{ss:reg}

In the Monte Carlo simulation, some dressed cluster states $\ket{R}^{\ell m_\ell}_{n_t}$ might become almost linearly dependent due to not being sampled enough, i.e., the simulation cannot fully resolve these states. This introduces some ill-conditioning in the norm matrix. It may be necessary to adopt methods to stabilize the calculation. In this work, we approached this issue in two ways: first, with the so-called \enquote{Ridge regression} or \emph{Tikhonov} regularization \cite{tikhonov1977solutions,bhl137258,doi:10.1137/1.9780898719697} and second by projecting the Hamiltonian onto those modes in the norm matrix that have a well-defined physical behaviour. Both methods will be explained briefly in the following. For more details, we refer to the supplementary material of \cite{Sarkar:2026wev}, where a similar study was performed for $\alpha$-$\alpha$ scattering at N${}^3$LO.

Since the norm matrix is Hermitian and positive semidefinite in the absence of statistical and numerical errors, it admits the spectral decomposition
\begin{equation}
N = O D O^\dagger,
\quad
O^\dagger O = O O^\dagger = \mathcal{I},
\quad
D=\operatorname{diag}
\left(\lambda_1,\lambda_2,\ldots,\lambda_M\right).
\label{eq:norm_spectral}
\end{equation}
We order the eigenvalues according to
\begin{equation}
\lambda_1\geq\lambda_2\geq\cdots\geq\lambda_M,
\label{eq:norm_ordering}
\end{equation}
so that the final entries of $D$ correspond to the least well-resolved directions of the norm matrix. 

\emph{Tikhonov regularization:}
We replace the norm matrix by
\begin{equation}
N_\kappa=N+\kappa A_n,
\end{equation}
where $\kappa>0$ is the regularization parameter and $A_n$ is the projector onto the subspace spanned by the $n$ eigenvectors with the smallest eigenvalues,
\begin{equation}\label{eq:A_O}
    A = O\, \mqty(0 & 0 \\ 0 & \mathcal{I}_{n\times n})\, O^{\dagger},
\end{equation}
Consequently, the addition of $\kappa$ prevents the inverse square root
$N_\kappa^{-1/2}$ from amplifying the least well-resolved norm
eigenmodes. We calculate the phase shifts for several values of
$\kappa$ and obtain the final result by extrapolating to
$\kappa\rightarrow0$. We verified that, within the range of
regularization parameters used in the analysis, the extrapolated
results are insensitive to the number $n$ of regularized modes.
We use $n=1$ in the final analysis so that the norm matrix is modified
in the smallest possible subspace.

\emph{Norm-mode projection:}
The eigenvectors of the norm matrix define the norm modes
$\ket{\eta_\alpha}$, with $\alpha=1,\ldots,M$. In terms of the radial cluster states introduced in \cref{eq:cluster_states}, they are
given by
\begin{equation}
\ket{\eta_\alpha}
=
\sum_{i=1}^{M}
O_{i\alpha}\ket{R_i},
\label{eq:norm_mode}
\end{equation}
where $O$ is the matrix defined in \cref{eq:A_O}. The norm modes satisfy
\begin{equation}
\braket{\eta_\alpha}{\eta_\beta}
=
\lambda_\alpha\delta_{\alpha\beta}.
\end{equation}
A small eigenvalue therefore corresponds to a nearly null linear
combination of radial cluster states. We order the eigenvalues according to
\begin{equation}
\lambda_1\geq\lambda_2\geq\cdots\geq\lambda_M
\end{equation}
and retain the set
\begin{equation}
K
=
\left\{
\alpha:
\lambda_\alpha>\epsilon\lambda_1
\right\},
\end{equation}
where $\epsilon$ is a dimensionless cutoff parameter. Let $O_K$ denote the $M\times M_K$ matrix formed by the retained eigenvectors, and let $D_K$ denote the corresponding $M_K\times M_K$ diagonal matrix of retained eigenvalues. The inverse square root of the norm matrix projected onto the \enquote{good} modes is then 
\begin{equation}
    X_\epsilon = O_K^{~} D_K^{-1/2} O_K^{\dagger}.
\end{equation}
Then, the projected adiabatic Hamiltonian is,
\begin{equation}
    H_\mathrm{adiabatic}^{(\epsilon)} = X_\epsilon H X_\epsilon.
\end{equation}
Both $X_\epsilon$ and $H_\mathrm{adiabatic}^{(\epsilon)}$ are $M\times M$ matrices whose row and column indices correspond to the original radial bins. Thus, no radial point is removed. The projection removes only the nearly null linear combinations of radial states associated with the discarded norm modes. Retaining the full radial-coordinate representation is important for connecting the interacting adiabatic Hamiltonian to the large-volume asymptotic Hamiltonian and for analyzing its eigenfunctions as functions of the radial separation.
We verified that the final phase shifts are stable under variations of $\epsilon$.

To access small relative momenta without performing the complete four-nucleon Monte Carlo calculation in a very large volume, we follow the finite-to-large-volume extension developed for radial adiabatic transfer matrices in Ref.~\cite{Elhatisari:2016hby}. In the asymptotic region, where the two cluster wave packets no longer overlap and the short-range nuclear interaction is negligible, the radial adiabatic matrix approaches an effective two-cluster matrix containing the free relative motion and any long-range interactions. We apply the corresponding construction here to the adiabatic Hamiltonian. For the projected $d+d$, ${}^{5}S_2$ channel, the asymptotic Hamiltonian describes two deuterons interacting only through the
Coulomb potential. We therefore connect the interacting adiabatic Hamiltonian calculated in the smaller volume to the two-deuteron Coulomb Hamiltonian and extend it to lattices as large as $L_{\mathrm{big}}=\SI{92}{\femto\m}$. We have verified that the extracted phase shifts are stable under variations of the connection radius, provided that it lies beyond the range of the short-range nuclear interaction. Nevertheless, we included possible differences of the phase shift with respect to different connection radii as a systematic error in our calculation.

\subsection{Calculating phase shifts}\label{s:phase_shifts}

\begin{figure*}[t!]
    \centering
    \resizebox{0.75\textwidth}{!}{\input{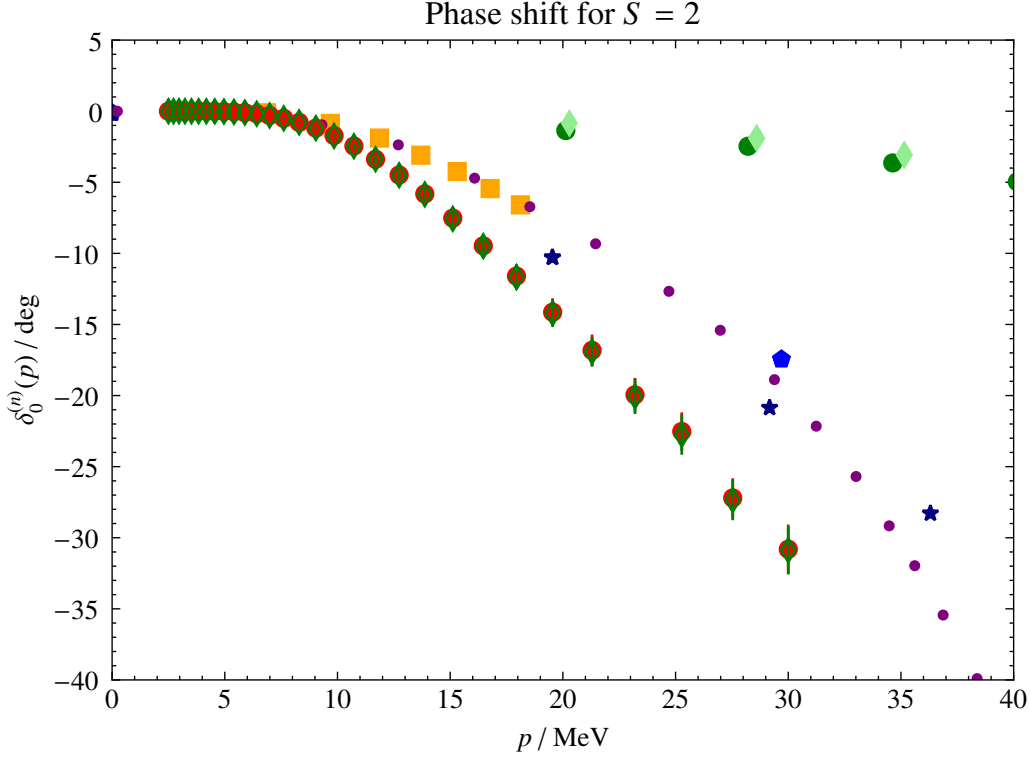}}
    \caption{
    Coulomb-subtracted phase shift of the projected single-channel $d+d$ elastic scattering phase shift for the ${}^{5}S_2$ channel from NLEFT at N$^3$LO in the chiral expansion with the Tikhonov regularization (green empty diamonds) and with the norm mode projection method (red circles) compared to results from the literature using 
    different continuum methods. The orange squares are results from a RGM calculation with a Malfliet-Tjon potential, the purple dots are the results presented in \cite{Filikhin:2000} from cluster-reduction method for solving the Yakubovskii equations with the same potential. The blue stars and pentagons are the phase shifts obtained in \cite{Hofmann:1996jv} and \cite{Hofmann:2005iy} from a R-matrix analysis of Helium-4 data, with a Bonn-based potential in \cite{Hofmann:1996jv} and the Argonne v18 two-nucleon plus Urbana IX three-nucleon potentials in \cite{Hofmann:2005iy}. The light-green diamonds and green circles are the most recent results presented in \cite{Carew:2021jai} from a variational-bound method.}
    \label{fig:phase_result}
\end{figure*}

To extract the phase shifts from the adiabatic Hamiltonian, we follow
the auxiliary-potential strategy of Ref.~\cite{Lu:2015riz}. We
supplement the radial Hamiltonian with a real auxiliary potential, 
localized near the outer radial boundary, given by,
\begin{equation}
    V_\mathrm{aux}(r) = C \cdot \exp\left\{-\frac{1}{2}
    \left(
\frac{r-R_{\mathrm{wall}}}{w_{\mathrm{aux}}}
\right)^2
\right\}\,.
\end{equation}
The width $w_{\mathrm{aux}}$ is chosen such that there exists a radial fitting interval between the short-range nuclear-interaction region and the auxiliary-potential region. In this interval, both the short-range nuclear interaction and $V_{\mathrm{aux}}(r)$ are negligible, while the Coulomb interaction remains present. Varying
$C$ shifts the discrete energy eigenvalues continuously and thereby allows us to determine phase shifts over a range of small relative momenta. In this region, the wave function takes the form
\begin{equation}\label{eq:wf}
    \psi(r) = A H_\ell^{-}(\eta,pr) - B H_\ell^{+}(\eta,pr),
\end{equation}
where $p$ is the momentum related to the energy eigenvalue $E$ of the Hamiltonian through the dispersion relation $p = \sqrt{2\mu E}$. $\mu = M_d/2$ is the reduced mass of the system with two scattering deuterons of mass $M_d$ and $\eta = \alpha_\mathrm{EM}\mu/k$ is the {Sommerfeld parameter}, with $\alpha_\mathrm{EM}$ the electromagnetic fine-structure constant. The functions $H_\ell^{\pm}$ are linear combinations of the Coulomb functions:
\begin{equation}
    H^{\pm}_\ell = G_\ell(\eta,pr) \pm i F_\ell(\eta,pr).
\end{equation}
Since the asymptotic channel space contains only the $d+d$, ${}^{5}S_2$ channel, its scattering matrix is one dimensional and reduces to a phase. The incoming and outgoing amplitudes are therefore related by
\begin{equation}
    B = A e^{2i\delta^{(n)}_\ell(p)},
\end{equation}
where $\delta^{(n)}_\ell(p)$ denotes the Coulomb-subtracted
phase shift of the single-channel $d+d$, ${}^{5}S_2$ Hamiltonian.
Fitting the eigenfunction of the adiabatic Hamiltonian to
\cref{eq:wf} determines this phase shift. We have assumed here that the effect coming from contributions to the wave function from the other $J^\pi = 2^+$ channels is negligible and that the $S$-matrix is unitary, i.e.\,can be written as $\exp(2i\delta)$, with $\delta$ denoting the total (real) phase shift.

We extract the phase shifts for all $L_t$ individually and then extrapolate to infinite Euclidean time. Details for this extrapolation are given in  \cref{app:ecl_time}.

\section{Results and discussion}\label{s:results}

For the results shown in the following we have used about $200 000$ GPU hours on JUWELS and JUPITER at Forschungszentrum J\"ulich. For each value of $L_t$, we perform several independent Monte Carlo simulations using different random-number seeds. We treat each independently seeded run as one statistical block and estimate the uncertainties using a delete-one-run jackknife. In each jackknife sample, one complete Monte Carlo run is omitted and the full analysis, including the construction and regularization of the adiabatic Hamiltonian and the extraction of the phase shifts, is repeated using the remaining runs. The resulting jackknife variance therefore includes the run-to-run fluctuations and propagates them through the
nonlinear analysis. The statistical uncertainty estimated within each individual simulation is much smaller than the variation among the
independent runs.

Our results for the ${}^5S_2$  phase shift are shown in \cref{fig:phase_result}. First, the results obtained with the two norm-matrix regularization methods described in \cref{ss:reg} agree within their uncertainties. Second, our phase shifts become more negative at increasing momentum than those reported in Refs.~\cite{Meier:1975ydi,Hofmann:1996jv,Filikhin:2000,Hofmann:2005iy,Carew:2021jai} corresponding to stronger effective repulsion. The three-nucleon interactions have only a small
effect, and omitting them makes the phase shifts slightly less negative. 

The origin of the differences between our results and those in the previous literature cannot be identified unambiguously. As also discussed in Ref.~\cite{Carew:2021jai}, differences in the nuclear interactions and numerical methods are likely to play a role. In addition, the calculations do not all employ the same channel space. Refs.~\cite{Hofmann:1996jv,Hofmann:2005iy} include explicit coupled-channel dynamics, whereas the remaining low-energy results shown in \cref{fig:phase_result} primarily correspond to projected $S$-wave $d+d$ calculations. Differences in the channel spaces may therefore contribute to the disagreement with the coupled-channel results, but cannot by themselves explain the discrepancy with all previous single-channel or $S$-wave-projected calculations. Other possible sources include differences in the two- and three-nucleon interactions, the treatment of Coulomb effects, and the numerical approximations employed in the different methods. The available results do not permit these effects to be separated quantitatively.

\begin{figure}[t!]
    \centering
    \includegraphics[width=\linewidth]{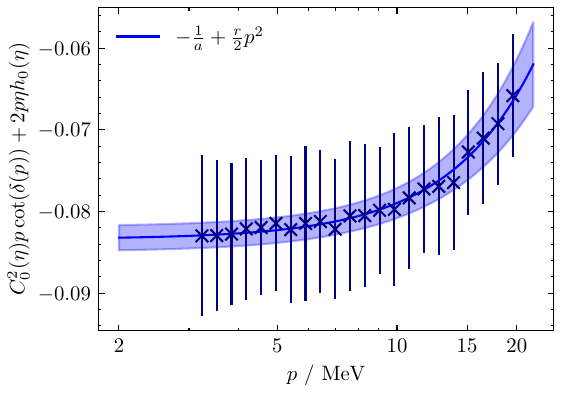}
    \caption{Fitting the modified effective range expansion (ERE) to \cref{eq:modERE}.}
    \label{fig:ERE}
\end{figure}

From the phase shifts we extract the Coulomb-modified scattering length and the effective range for the ${}^{5}S_2$ channel, ${}^{5}a_{dd}$ and ${}^{5}r_{dd}$, respectively, by fitting the modified effective-range
expansion
\begin{equation}\label{eq:modERE}
    C_0^2(\eta) p \cot(\delta_0^{(n)}) + 2p\eta h_0(\eta) = -\frac{1}{{}^5a_{dd}} + \frac{{}^5r_{dd}}{2}p^2 + {\cal O}(p^4)~, 
\end{equation}
with the Sommerfeld-Gamow factor
\begin{equation}
    C_0^2(\eta) = \frac{2\pi\eta}{\exp(2\pi\eta)-1}
\end{equation}
and 
\begin{equation}
    h_0(\eta) = -\log\abs{\eta} + \Re{\psi(i\eta)}~,
\end{equation}
where $\psi(z) = \Gamma'(z)/\Gamma(z)$ is the digamma function. From the fit displayed in \cref{fig:ERE}, we find for the scattering length and effective range
\begin{align}
    {}^5a_{dd} = \SI{11.99 \pm 0.22}{\femto\m}, \label{eq:scatt_length} \\
    {}^5r_{dd} = \SI{3.44 \pm 0.79}{\femto\m}.
\end{align}

The scattering length is much larger than the value of ${}^5a_{dd} = \SI{7.8\pm 0.3}{\femto\m}$ quoted in \cite{Carew:2021jai} or the ${}^5a_{dd} = \SI{7.5}{\femto\m}$ from \cite{Filikhin:2000}, and about twice the value of the scattering length found in \cite{Rojik:2025jcv}, which is ${}^5a_{dd} = \SI{6.26 \pm 0.03}{\femto\m}$. However, in Ref.~\cite{Rojik:2025jcv} the NLO correction to the LO results is about 50\%. We have checked the consistency of our calculation by performing the fit with an additional $v_2 p^4$ term in the effective range expansion and by fitting only a constant to the data up until $\SI{5}{MeV}$. In both cases the obtained scattering length was within uncertainties of the result given in \cref{eq:scatt_length}. The lack of data points for small momenta given in \cite{Filikhin:2000} and \cite{Carew:2021jai} makes it difficult to reproduce their result for the scattering length or, stated differently, makes a quantitative guess on the origin of the differences to our results impossible. 

We also find that our results for the phase shift are consistent with a calculation on a larger lattice with $L = \SI{21.1}{\femto\m}$, which would mean that finite-volume effects are negligible, however, the uncertainties for this calculation are quite large. In order to refine the calculations presented in this work, one would need much larger computational resources in order to get a proper quantitative estimate on the finite-volume effects. 

The present calculation establishes the feasibility of treating the $d+d$, $^{5}S_2$ channel with NLEFT and identifies the numerical requirements for extending the calculation to coupled reaction channels. It will certainly be of interest to calculate the cross sections for the deuteron-deuteron reactions relevant for BBN  within NLEFT and so a full coupled-channel analysis including the  $d+d$, $^{1}S_0$ and $^{5}D_0$, as well as the $p+{}^3\mathrm{H}$ and $n+{}^3\mathrm{He}$ $^{1}S_0$ channels contributing to the $J^\pi = 0^+$ system is currently underway.

\section*{Acknowledgements}
We are grateful to the members of the NLEFT collaboration, in particular Avik Sarkar for support and useful discussions.
This work was supported in part by the European
Research Council (ERC) under the European Union's Horizon 2020 research
and innovation programme (EXOTIC, grant agreement No. 101018170),
and by the CAS President's International Fellowship Initiative (PIFI) (Grant No.~2025PD0022).  The authors gratefully acknowledge the Gauss Centre for Supercomputing e.V. (www.gauss-centre.eu)
for funding this project by providing computing time on the GCS Supercomputers JUWELS and JUPITER
at J\"ulich Supercomputing Centre (JSC) and the support of the project \mbox{EXOTIC} by the JSC by dedicated HPC time provided on the \mbox{JURECA DC} GPU partition. Furthermore, the authors gratefully acknowledge the computing time provided on the high-performance computer
HoreKa by the National High-Performance Computing Center at KIT (NHR@KIT). This center is
jointly supported by the Federal Ministry of Education and Research and the Ministry of Science,
Research and the Arts of Baden-Württemberg, as part of the National High-Performance Computing
(NHR) joint funding program (https://www.nhr-verein.de/en/our-partners). HoreKa is partly funded
by the German Research Foundation (DFG).

\bibliographystyle{elsarticle-num} 
\bibliography{refs}

\begin{onecolumn}
\appendix
\section{Euclidean time extrapolation of phase shift results}\label{app:ecl_time}
\setcounter{figure}{1}
For each individual momentum point we calculate the phase shift as described in \cref{s:phase_shifts} for $L_t = 20, 30, 40, 50$ and $60$. We then fit an exponential
\begin{equation}
    \delta_0 (p ; L_t) = \delta_\infty(p) + c(p) \exp(-\Delta E(p) L_t \alpha_t),
\end{equation}
so $\delta_\infty(p)$ is the phase shift in the limit of $L_t\to\infty$. Here, we constrain the fits in such a way, that $\Delta E$ is the same for five neighbouring momentum points. The fits for 15 momentum points are shown in \cref{fig:Ecltime_extr_projection,fig:Ecltime_extr_tikhonov} for the case where we used the norm mode projection method and Tikhonov regularization, respectively. In most figures the fit appears as a constant: it seems that we have reached a plateau in the exponential already at $L_t=20$. For the sake of focussing on the results obtained from the MC simulation, we show the fit result only in the range where we have data points.

\bigskip

\begin{figure}[h!]
    \centering
    \includegraphics[width=\textwidth]{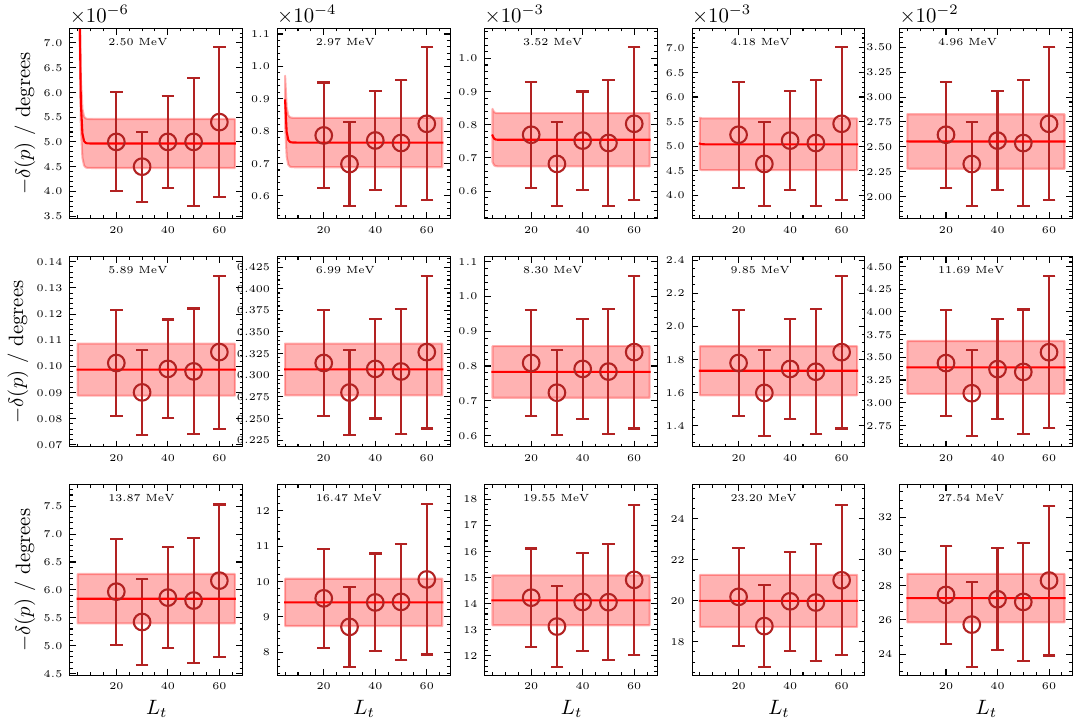}
    \caption{Euclidean time extrapolation for the $d+d$ phase shift calculated with the auxiliary potential and norm mode projection method (red circles). The red line and band give the results from a bootstrap fit of an exponential function to the data.}
    \label{fig:Ecltime_extr_projection}
\end{figure}

\begin{figure}
    \centering
    \includegraphics[width=\textwidth]{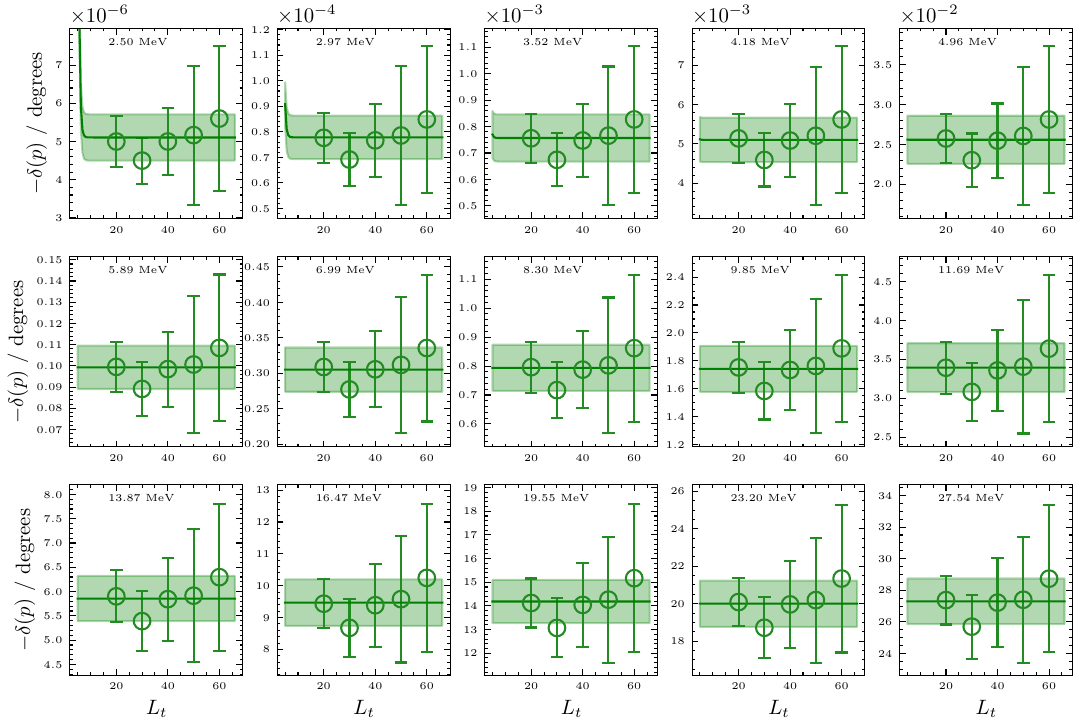}
    \caption{Euclidean time extrapolation for the $d+d$ phase shift calculated with the auxiliary potential method and Tikhonov regularization (green circles). The green line and band give the results from a bootstrap fit of an exponential function to the data.}
    \label{fig:Ecltime_extr_tikhonov}
\end{figure}

\end{onecolumn}

\end{document}